\documentclass[%
reprint,
superscriptaddress,
nofootinbib,
amsmath,amssymb,
aps,
pra,
notitlepage,
showkeys]{revtex4-2}

\usepackage{color}
\usepackage{url}
\usepackage{graphicx}
\usepackage{lipsum}
\usepackage[export]{adjustbox}
\usepackage[font=footnotesize,labelfont=bf]{caption}
\usepackage{mathtools}
\usepackage{gensymb}
\usepackage[T1]{fontenc}
\usepackage{amsmath}
\usepackage{amssymb}
\usepackage{hyperref}
\usepackage{stmaryrd}
\usepackage{fancyhdr}
\usepackage[utf8]{inputenc}
\usepackage{rotating}
\usepackage{comment}
\usepackage{url}
\usepackage{bbm}
\usepackage{soul}
\usepackage[normalem]{ulem}
\usepackage{tabularx,booktabs}
\usepackage{stmaryrd}
\newcolumntype{Y}{>{\centering\arraybackslash}X}

\newcommand{\iu}{\mathrm{i}\mkern1mu}
\newcommand{\dint}{\mathrm{d}}

\begin{document}
\title{Exact time-dependent dynamics of discrete binary choice models} 
\author{James Holehouse}
\email{james.holehouse@ed.ac.uk}
\affiliation{School of Biological Sciences, University of Edinburgh, Rm 3.04, CH Waddington Building, Max Born Crescent, Edinburgh, EH9 3BF}

\author{Jos\'{e} Moran}
\affiliation{Mathematical Institute and Institute for New Economic Thinking at the Oxford Martin School, University of Oxford, Oxford, United Kingdom}
\affiliation{Complexity Science Hub Vienna, Josefst\"adter Stra{\ss}e 39, A-1080, Austria}

\begin{abstract}
We provide a generic method to find full dynamical solutions to binary decision models with interactions. In these models, agents follow a stochastic evolution where they must choose between two possible choices by taking into account the choices of their peers. We illustrate our method by solving Kirman and Föllmer's ant recruitment model for any number $N$ of agents and for any choice of parameters, recovering past results found in the limit $N\to \infty$. We then solve extensions of the ant recruitment model for increasing asymmetry between the two choices. Finally, we provide an analytical time-dependent solution to the standard voter model and a semi-analytical solution to the vacillating voter model. 
\end{abstract}

\date{\today}
\maketitle

\section{Introduction}

That individuals take into account the choices made by others when making their own is evident to anyone who has witnessed fashion fads, trends and events of mass panic like bank runs. These collective phenomena have dramatic social consequences, as they are authentic ``collective delusions''~\cite{mackay2003extraordinary}, of which economic bubbles and the subsequent crashes they produce are eloquent examples. The mechanism through which they appear is intuitive: sociable individuals tend to imitate the choices made by their peers, choosing to go to the same restaurant, dress the same way or buy/sell the same asset as their group of friends or the collective zeitgeist dictates. For any of these, when a choice becomes that of the majority its dominance and attractiveness tends to increase, as more and more individuals are persuaded to make it.

To the physicist, this is reminiscent of the mechanism governing certain phase transitions, and in particular that of the ferromagnetic transition, where the magnetic dipoles in a material all suddenly point in the same direction when cooled below a critical temperature. Owing to the common points between these mechanisms, similar behaviour is observed in the abrupt opinion swings seen in certain social systems (see~\cite{Michard2005} and references therein).

Thus it is no surprise that one of the strongest criticisms to the old paradigm of the \textit{rational representative agent} used in textbook economics is that it does not sufficiently take into account interactions between agents. In that framework, agents make the choice that maximises a certain \textit{utility function}, quantifying the level of satisfaction procured by said choice, by taking into account the different constraints they face---such as a limited budget. 

Because there are no interactions, these models fail to capture the rich collective phenomena, or even the crises, that appear in real social systems~\cite{bouchaudcrises}. For example, in a system made of non-interacting rational agents the only explanation for a large opinion swing is an \textit{exogenous} event, such as the publication of new information that influences the agents. Therefore it is necessary to go further to understand the link between the \textit{micromotives} that guide agents and their collective \textit{macrobehaviour}~\cite{schelling2006micromotives}.

A number of efforts have been made to alleviate this issue, notably by considering models where agents' decisions are influenced by interactions with their peers~\cite{Schelling1971,brock2001discrete,kirman1993ants,bouchaudcrises,Redner2019}. These models often study cases where agents face only two possible choices---reducing the problem to that of making a binary decision. In this way, one can study \textit{toy models} describing social systems where agents interact, in the hope of gaining a better understanding of collective social phenomena much like the Ising model set a precedent for the understanding of emergent phenomena in condensed matter physics.

In spite of their simplicity, these models show a very rich phenomenology characterised by the appearance of crises, hysteresis and other emergent phenomena \cite{hosseiny2019hysteresis,bouchaudcrises}. However, these dynamical models have often been studied only once their \textit{stationary} state is reached, focusing in how their statistical description can change radically through subtle variations of the parameters that define it. But more insight can be gained by studying the full dynamics of how said stationary state is actually reached.

Indeed, one can consider Kirman and Föllmer's seminal ant recruitment model~\cite{kirman1993ants}. In its origin, it focused in explaining the results of an entomological experiment where an ant colony had access to two identical food sources. Instead of spreading evenly between the two sources, the ants were observed to concentrate in one of the two sources before randomly switching collectively to the other~\cite{Deneubourg1990}. Similar models have arisen in independent parts of the literature---indeed the ant rationality model is very similar to the Bass diffusion model \cite{bass1969new,young2009innovation}, which describes the uptake of a new product or practice in a population, and if modelled stochastically would lead one a model isomorphic to the model of Kirman and Föllmer.

The authors of~\cite{kirman1993ants} showed that this could be understood through a model were the ants had a certain propensity to imitate their peers, and another propensity to switch randomly between the two sources. When the effect of imitation is strong, the distribution of the number of ants in the food sources is bi-modal, and so one is more likely to find a majority of ants in either of the two sources, while when the random switching dominates one finds a regime with an unimodal distribution, with a rough half-and-half split between the two sources.

Although inspired by an example coming from behavioural biology, this model, and others that are very similar, has been used to explain behaviour in financial markets~\cite{Alfarano2007,alfarano005,Sano2014}, firm agglomeration~\cite{dindo_bottazzi}, the dynamics of fishing boats~\cite{Moran2021} and even wealth inequality~\cite{bouchaud2021selffulfilling}. The model is in fact also identical to the Moran model in genetics~\cite{moran1958random}, and is also closely related to the Pólya urn model reviewed in~\cite{Pemantle2007}.

An interesting aspect of this model was found in~\cite{moran2020schrodinger}, where a full dynamical solution to the model was provided in the limit of an infinitely large number of ants. Indeed, a key finding is that the time it takes for the ant colony to switch collectively from one food source to the other depends \textit{exclusively} on the rate at which ants switch randomly. This can then be interpreted as implying that collective switches are driven by a single ant going to the other source and attracting all the others through an imitative avalanche. It is therefore clear that a precise dynamical description of such models is key in understanding the collective behaviours they display.

In this article, we solve this model in the case of a finite number of ants and show how the results from~\cite{moran2020schrodinger} can be recovered. We also show how our methods can be extended to solve a large class of similar models, such as the voter model~\cite{Redner2019,lambiotte2007dynamics}. 

The paper is structured as follows. In the first section we describe the ant recruitment model fully, and show how to map it onto a birth/death process. We solve it analytically using generating functions, and also obtain semi-analytical results in a computationally efficient way using the methods described in~\cite{smith2015general}. We then apply these methods to solving a more general version of the model, taking into account all possible asymmetries. Finally, we show applications of these techniques to the voter and vacillating voter models.

\section{Setup}

We first illustrate our setup by solving the stochastic dynamics of Kirman and Föllmer's ant rationality model. Consider a system of $N$ ants where there are two different sources of food, \textit{left} $L$ and \textit{right} $R$. Each ant is associated with a single food source, and we denote $n$ as the number of ants at the right-hand food source. Since we do not track the spatial position of the ants, $n$ completely specifies the state of the system.

The ants are subject to two separate influences: (i) a random influence whereby \textit{each ant switches to the opposite food source at rate $\varepsilon$}, and (ii) a collective influence whereby when two ants meet---at rate $\nu$---\textit{if they are associated with opposing food sources, then one of the ants recruits the other to its food source.} Given that any two ants meet at rate $\nu$ \textit{regardless of their current food source}, it is straightforward to show that the propensity at which two ants at opposing food sources meet is $\tilde{\nu}(n)=n(N-n)\nu/(N-1)$. We can now write a dynamical effective reaction scheme describing the number of ants on the right hand food source,
\begin{align}\label{eq:kirmanReac}
    L \xrightleftharpoons[n\varepsilon+\tilde{\nu}(n)]{(N-n)\varepsilon +\tilde{\nu}(n)} R.
\end{align}
Note that unlike effective reaction schemes often written in chemical reaction networks~\cite{schnoerr2017approximation} the expressions labelling the arrows denote the full propensity for the event to occur given the state of the system $n$.

From this effective reaction scheme one can then describe the dynamical evolution of the probability distribution $P(n,t)$ for reaction scheme \eqref{eq:kirmanReac} via the following master equation,
\begin{widetext}
\begin{equation}\label{eq:MEKirman}
\begin{split}
\partial_t P(n,t) = &\left[(N-(n-1))\varepsilon +\tilde{\nu}(n-1)\right]P(n-1,t)+ \left[(n+1)\varepsilon+\tilde{\nu}(n+1)\right]P(n+1,t)\\
&-\left[(N-n)\varepsilon +n\varepsilon+2\tilde{\nu}(n)\right]P(n,t),
\end{split}
\end{equation}
\end{widetext}
with a given initial condition that $n(t=0)=n_0$ ants are initially at the right-hand food source, represented by $P(n,0)=\delta_{n,n_0}$ where $\delta_{i,k}$ is the Kronecker delta symbol.

Defining the $(N+1)\times(N+1)$-dimensional real matrix $\mathbf{M}$ as
\begin{equation}\label{eq:M_def}
\begin{split}
\left(\mathbf{M}\right)_{n,m} = &\delta_{n-1,m}\left((N-(n-1))\varepsilon + \tilde{\nu}(n-1)\right)\\
&+ \delta_{n+1,m}\left((n+1)\varepsilon + \tilde{\nu}(n+1)\right)\\
&- \delta_{n,m}\left(N\varepsilon + 2\tilde{\nu}(n)\right),
\end{split}
\end{equation}
it is straightforward to see that the master equation can be re-cast as $\partial_t \vec{P}(t) = \mathbf{M}\vec{P}$, where the $n$-th element of $\vec{P}(t)$ is $P(n,t)$. The matrix $\mathbf{M}$ corresponds to the Liouville or master operator and completely describes the dynamics of our system as it contains all the information on the transition rates.

The steady state distribution $\vec{P}_0$ can be derived by solving the equation $\mathbf{M}\vec{P}_0=0$.  This can be solved by recursion, taking afterwards the $N\to\infty$ limit, with $n/N=x$ fixed to that the stationary distribution is given by a symmetric Beta distribution~\cite{kirman1993ants,moran2020schrodinger}. That is,

\begin{equation}\label{eq:stationary}
P_s(n)\equiv P(n,t\to\infty) \underset{N\gg 1}{\propto} \left(\frac{n}{N}\right)^{\varepsilon/\mu-1}\left(1-\frac{n}{N}\right)^{\varepsilon/\mu-1},
\end{equation}
where we  introduce $\mu=\frac{\nu}{N-1}$ so that $\tilde{\nu}(n) = n(N-n)\mu $ to match the notation of~\cite{moran2020schrodinger}.

\begin{figure}[tb]
    \centering
    \includegraphics[width=.5\textwidth]{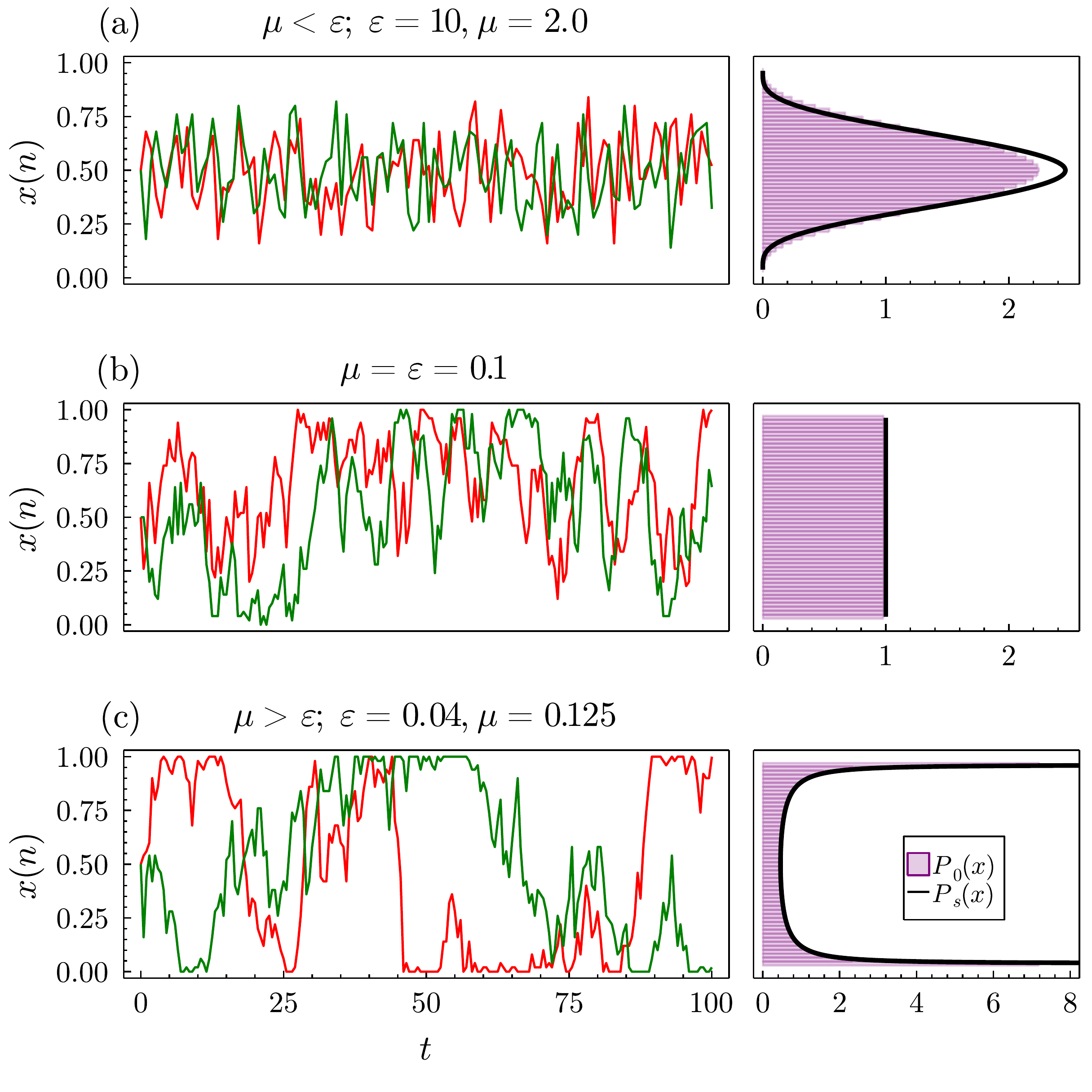}
    \caption{Sample of three different trajectories (left) of the fraction of ants $x(n)=n/N$ on the right-hand source, with the red and green lines showing two different realisations for $N=50$ ants, along with the corresponding stationary densities (right). Purple bars are the exact stationary distribution for finite $N$, $P_0(x)=NP_0(n)$, and the black lines correspond to the symmetric Beta distribution, $P_s(x)=N P_s(n)$, given in Eq.~\eqref{eq:stationary}. The stochastic simulations are done using the stochastic simulation algorithm \cite{gillespie2007stochastic}. Notice that in the high imitation regime $\varepsilon<\mu$, corresponding to plot (c) the ants tend to concentrate in one of the food sources for a time of order $1/\varepsilon$ before switching collectively to the other source. The case $\varepsilon=\mu$ in (b) corresponds to the situation where the Beta distribution is a uniform distribution over $[0,1]$.}
    \label{fig:ssa_sims}
\end{figure}

This is the main point of interest of this model, as stressed by Kirman~\cite{kirman1993ants}: when imitation is strong, with $\varepsilon<\mu$, the most probable state is to have all of the ants in a single food source, as shown by the divergence of the probability distribution in Eq.~\eqref{eq:stationary}, while the same probability is $\approx 0$ in the high-noise regime $\varepsilon>\mu$ where the most probable state is to have a $50/50$ split between the two sources. We exhibit this behaviour in Fig.~\ref{fig:ssa_sims}, and show that even for $N=50$ the Beta distribution is a good approximation to the exact distribution. Note that where $\varepsilon/\mu=1$ we get uniform distributions for both $P_0(n)$ and $P_s(n)$

Because the tri-diagonal coefficients $(n,n+1)$ and $(n+1,n)$ are positive, and because the rank of the matrix is clearly $N+1$, it is straightforward to show that $\mathbf{M}$ is composed of $N+1$ distinct \textit{real} eigenvalues that we label $-\lambda_m$ for $m=0,\ldots,N$.\footnote{Indeed, if one considers $\mathbf{J}=\mathbf{D}^{-1}\mathbf{MD}$, with $\mathbf{D}=\text{diag}(M_{n,n+1}M_{n+1,n})$ it is easy to show that $\mathbf{J}$ is symmetric. $\mathbf{M}$ is therefore similar to a symmetric matrix, and has real eigenvalues.} Further, direct application of the Perron-Frobenius theorem shows that $\lambda_m\geq 0$ and that $0$ is an eigenvalue of $\mathbf{M}$. We choose therefore to label these eigenvalues as $0=\lambda_0<\lambda_2<\ldots<\lambda_{N}$. 

The model may now be formally solved as $\vec{P}(t) = e^{t\mathbf{M}}\vec{P}(0)$. We may then write $\mathbf{M} = -\sum_m \lambda_m \vec{U}_m \vec{V}_m^T$,\footnote{Indeed, write $\mathbf{M} = - \mathbf{P\Delta P}^{-1}$ with $\Delta = \text{diag}(\lambda_m)$, then $(\vec{U}_m)_i = P_{im}$ and $(\vec{V}_m)_i = (\mathbf{P}^{-1})_{im}$. These vectors are known respectively as the right- and left-eigenvectors of $\mathbf{M}$.} which leads to $\vec{P}(t) = \sum_m e^{-\lambda_m t}\left(\vec{V}_m\cdot \vec{P}(0)\right)\vec{U}_m$. Denoting finally $c_m = (\vec{V}_m \cdot \vec{P}(0))$ and $\vec{U}_m = \left(f_m(0),\ldots, f_m(N)\right)^T$ we reach the following formula for the full solution:

\begin{equation}\label{eq:formal_sol}
P(n,t) = \sum_m c_m f_m(n) e^{-\lambda_m t},
\end{equation}
where the different terms $c_m, f_m(n)$ and $\lambda_m$ remain to be determined.

This is a formal solution of a discrete master equation. Master equations are notorious for being difficult to solve, \textit{especially in time}. Common methods include the Poisson representation~\cite{gardiner2009stochastic}, Fokker-Planck (or Langevin) approximations~\cite{van1992stochastic,gardiner2009stochastic,moran2020schrodinger,gillespie2000chemical}, field theory~\cite{tauber2014critical}, the linear-mapping approximation~\cite{cao2018linear} and 
the system-size expansion~\cite{van1992stochastic,thomas2014system,thomas2015approximate}. Below we utilise a combination of other methods, notably, the method of generating functions~\cite{van1992stochastic,gardiner2009stochastic}, eigenfunction methods~\cite{van1992stochastic,gardiner2009stochastic} and the time-dependent solution to the 1D master equation~\cite{smith2015general}.

In particular, the method used in~\cite{moran2020schrodinger} reached a solution of the same form as Eq.~\eqref{eq:formal_sol} by properly taking the limit $N\to\infty$ as to transform the matrix $\mathbf{M}$ into a Fokker-Planck partial-differential operator. The eigenfunctions (equivalent to the eigenvectors of $\mathbf{M}$) and eigenvalues of that operator were then found by two successive changes of variables mapping the problem onto a solvable quantum-mechanical problem. We claim to reach equivalent results using simpler methods that can be reused for other, similar models transparently.

\subsection{Explicit solution} 
\label{sub:explicit_solution}


The one-dimensional nature of the problem, along with the form of Eq.~\eqref{eq:formal_sol}, invites us to introduce the generating function $G(z,t)=\sum_n z^n P(n,t)$, defined for $\vert z \vert \leq 1$. Plugging this definition into the ordinary differential equation system of Eq.~\eqref{eq:MEKirman} we obtain the following partial differential equation,
\begin{equation}\label{eq:kirmanGF}
\begin{split}
\frac{\partial_t G(z,t)}{z-1}=&\varepsilon N G(z,t)+(\mu(N-1)(z-1)\\
&-\varepsilon (z+1))\partial_z G(z,t)\\
&-\mu z(z-1)\partial_z^2 G(z,t),
\end{split}
\end{equation}
where we denote again $\mu = \nu/(N-1)$. This generating function PDE is subject to a boundary condition and an initial condition; the boundary condition relates to the normalisation of probability and is $G(1,t)=1$, while the initial condition at $t=0$ is found to be $G(z,0) = \sum_n \delta_{n,n_0}z^n = z^{n_0}$. Note that probabilities and moments can be obtained directly from the generating function:
\begin{equation}\label{eq:moments}
\begin{split}
P(n,t) &= \frac{1}{n!}\partial_z^n G(z,t)|_{z=0},\\
\mathbb{E}[(n)_r] &= \partial_z^n G(z,t)|_{z=1},
\end{split}
\end{equation}
where $\mathbb{E}[(n)_r]= \mathbb{E}\left[\prod_{i=0}^{r-1}(n-i)\right]$ is the $r^{\mathrm{th}}$ factorial moment.

From Eq.~\eqref{eq:formal_sol} it is clear that this function can be written as $G(z,t)=\sum_m c_m \left(\sum_n f_m(n)z^n\right) e^{-\lambda_m t}$. Defining now $g_m(z)= \sum_n f_m(n) z^n$ we reach the same form we would have obtained had we used an exponential ansatz for the solution~\cite{van1992stochastic,tauber2014critical}, namely $G(z,t) = \sum_m c_m g_m(z)e^{-\lambda_m t}$. The interpretation of the $g_m(z)$ functions is transparent, as they are the ``generating functions'' associated to the pseudo-distributions $f_m(n)$.\footnote{Note that this is an abuse of language, as only $f_0(n)$ is proportional to an actual probability distribution as it corresponds to the steady state.}

This is the same ansatz used in time-dependent solutions to quantum mechanical problems~\cite{weinberg1995quantum}, where it arises naturally from the separation of variables $G(z,t)=f_1(z)f_2(t)$. Note that $g_0(z)$ corresponds, up to a normalisation constant, to the generating function of the steady state distribution $P(n,t\to\infty)\propto f_0(n)$.

Plugging this into Eq.~\eqref{eq:kirmanGF} we reach an ODE in terms of $z$ alone,
\begin{widetext}
\begin{equation}\label{eq:GFE}
\mu z(z-1) g_m''(z) - (\mu(N-1)(z-1)-\varepsilon(z+1))g_m'(z) -\left( \frac{\lambda_m}{z-1}+\varepsilon N \right)g_m(z)=0.
\end{equation}
\end{widetext}

 One finds the singularities of this ODE are at $z=0,1 \text{ and }\infty$ and are regular, hence the solution for $g_m(z)$ is given by a sum of two linearly independent hypergeometric type basis functions,
\begin{widetext}
\begin{equation}\label{eq:gzkirman}
\begin{split}
g_m(z) = (z-1)^{\alpha_m}&\Big\{c^{(m)}_1 \; {}_2F_1\left(\alpha_m+\frac{\varepsilon}{\mu},\alpha_m-N;1-N-\frac{\varepsilon}{\mu},z\right)\\
&+c^{(m)}_2 \; z^{N+\frac{\varepsilon}{\mu}} {}_2F_1\left(\alpha_m+\frac{\varepsilon}{\mu},\alpha_m+N;1+N+\frac{\varepsilon}{\mu},z\right)
\Big\},
\end{split}
\end{equation}
\end{widetext}
where 
\begin{equation}\label{eq:alpha_def}
    \alpha_m = \frac{\mu-2\varepsilon+\sqrt{4\varepsilon^2-4\varepsilon\mu+4\lambda_m\mu+\mu^2}}{2\mu}.
\end{equation}

However, owing to the definition of the generating function $G(z,t)$, the functions $g_m(z)$ should be polynomials of degree $N$ in $z$. We recall the definition of the hypergeometric function,
\begin{equation}\label{eq:2f1def}
_2F_1(a,b;c,z) = \sum_{\ell=0}^{\infty} \frac{(a)_\ell (b)_\ell}{(c)_\ell}z^{\ell}
\end{equation}
where $(a)_{\ell}=\prod_{j=0}^{\ell} (a+j)$ is the Pochhammer symbol. From this definition, one can check that this function is a polynomial only when either the first or second argument is a negative integer. This must hold for all possible values of $\varepsilon/\mu$, which means that $\alpha_m-N$ or $\alpha_m+N$ should be negative integers.

Consider now the first term in Eq.~\eqref{eq:gzkirman}: if $\alpha_m -N$ is a negative integer, i.e. $\alpha_m \in \llbracket 0,N\rrbracket$, we have a polynomial of degree $N-\alpha_m$ for the hypergeometric function which becomes a polynomial of degree $N$ after multiplication with $(z-1)^{\alpha_m}$, as required. We now relabel $c_1^{(m)}=c_m$.

On the other hand, for the second term we should have that $\alpha_m+N$ is a negative integer, suggesting to take an integer $\alpha_m \leq -N$ and giving a polynomial of degree $(-\alpha_m)-N$ for the hypergeometric term. However this is multiplied afterwards by $(z-1)^{\alpha_m}$, and one does not obtain a polynomial but a rational function. Therefore the only admissible solutions have $c_{2}^{(m)}=0$. 

We can therefore only keep the first term in the right-hand side of Eq.~\eqref{eq:gzkirman} and identify $\alpha_m$ with the index $m\in \llbracket 0;N\rrbracket$, allowing us to find the $N+1$ eigenvalues of our problem,
\begin{equation}\label{eq:eigvals}
 \lambda_m=m\left(2\varepsilon+(m-1)\mu\right), \quad m\in \llbracket 0; N\rrbracket,
 \end{equation} 
which are precisely those given in~\cite{moran2020schrodinger}, with the caveat that here $\mu$ depends explicitly on $N$ as $\mu = \nu/(N-1)$. Without loss of generality, we set $c_1^{(m)}=1$ and absorb it into the definition of $c_m$. 

The constant $c_m$ can then be evaluated by projecting the initial condition $G(z,0)=z^{n_0}$ onto the eigenfunctions $g_m(z)$, which form an orthogonal eigenbasis for a certain scalar product that can be determined fully using Sturm-Liouville theory (see Appendix~\ref{ap:sl_theory}). In other words, there exists a function $w(z)$ such that $\langle g_m, g_n\rangle = \int_{-1}^{1}\dint z~w(z) g_m(z) g_n(z) ) = \delta_{m,n}$. It follows then that 
\begin{equation}\label{eq:cm_def}
c_m = \frac{\int_{-1}^1 \dint z~w(z) z^{n_0}g_m(z)}{\int_{-1}^{1}\dint z~w(z)(g_m(z))^2},
\end{equation}
which is equivalent to the projection method on the orthogonal eigenfunctions of the imaginary-time Hamiltonian used in~\cite{moran2020schrodinger}.

We attract the reader's attention to the fact that the second eigenvalue $\lambda_1$ is still independent of $N$ and equal to $2\varepsilon$: the convergence to the stationary state, and therefore the rate at which ants switch to another source, is proportional only to the random switching rate, as found in~\cite{moran2020schrodinger} in the large $N$ limit. We note that the waiting time to switch between the two food sources was explored more in depth in \cite{biancalani2014noise}, where they approximately found the mean time it takes an ant to switch food sources for a given $(\varepsilon,\mu)$ as $N\to\infty$ based on first passage time theory.

It is also possible to retrieve the result from~\cite{moran2020schrodinger} that $\mathbb{E}[n(t)]-\mathbb{E}[n(t\to\infty)]\propto e^{-2\varepsilon t}$. Starting from the second line of Eq.~\eqref{eq:moments}, we write $\mathbb{E}[n(t)]=\sum_m c_m g'_m(1)e^{-\lambda_m t}$. Owing to the term $(z-1)^m$ in $g_m(z)$, it is quite straightforward to show that $g'_m(1)=0$ for $ m\geq 2$. Therefore we obtain that $\mathbb{E}[n(t)]= c_0g_0'(1)+c_1 g_1'(1) e^{-2\varepsilon t}$ as required, with $\mathbb{E}[n(t\to\infty)] = c_0 g'_0(1)=1/2$ because of symmetry considerations. 

Note also that the spectrum obtained in Eq.~\eqref{eq:eigvals} matches that of the $\tan^2$ P\"oschl-Teller potential~\cite{Nieto} for the quantum problem solved in~\cite{tacseli2003exact}. The corresponding Schr\"odinger's equation is solved by a trigonometric change of variables that puts the eigenvalue problem into the form of an Euler hypergeometric differential equation, similar to the one obtained in Eq.~\eqref{eq:GFE}. The discrete eigenvalues are then found by imposing that the wave-function must be square-normalisable, much as we must impose that the generating function be a polynomial in $z$. 

The method for the large $N$ limit used in~\cite{moran2020schrodinger} mapped the ant model into the $\tan^2$-potential Schrödinger's equation by writing a Fokker-Planck equation describing the random dynamics of the variable $x=n/N$, changing variables into $\varphi=2x-1$ to obtain another Fokker-Planck equation with a diffusive term that did not depend on $\varphi$ and finally by using another common technique, described in detail in~\cite{risken1996fokker}, to map this equation into a Schr\"odinger's equation. The method shown above achieves the same result in a much more straightforward way that can be applied to other similar problems and that allows one to obtain a solution for any value of $N$.

\subsection{Practical evaluation of $P(n,t)$} 
\label{sub:practical_evaluation_of_}
\begin{figure}[tb]
    \centering
    \includegraphics[width=.4\textwidth]{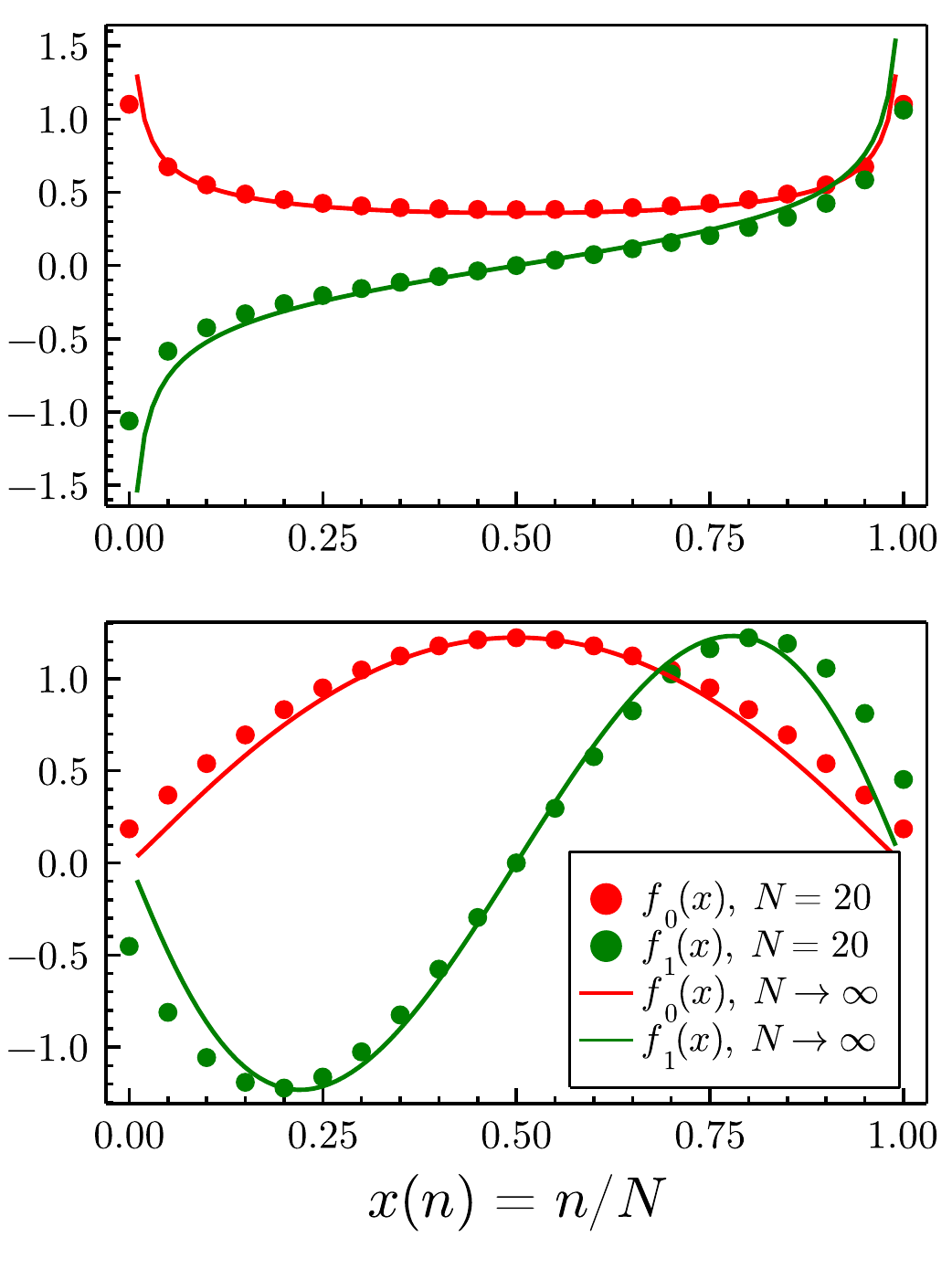}
    \caption{A figure showing the two first modes $f_0$ and $f_1$, represented with the reduced variable $x=n/N$. The solid lines correspond to the $N\to\infty$ results from~\cite{moran2020schrodinger} for $\varepsilon/\mu=0.6$ (left) and $\varepsilon/\mu=2.1$ (right), while the dots represent their discrete equivalent using Eq.~\eqref{eq:sol_fn} with $N=20$. The fit is already remarkably good at only $N=20$. In line with the quantum-mechanical interpretation given in~\cite{moran2020schrodinger}, the mode $f_1$ can be interpreted as describing the hopping of ants from one source to the other, hence its asymmetric shape about $x=0.5$. Note that this Figure reproduces Figure 2 from~\cite{moran2020schrodinger}.}
    \label{fig:solutions}
\end{figure}

Using the polynomial expression expressed above, ${}_2F_1(-k, a;b,z)=\sum_{l=0}^k \binom{k}{\ell}(-1)^\ell \frac{\Gamma(a+\ell)}{\Gamma(a)}\frac{\Gamma(b)}{\Gamma(b+\ell)}z^\ell$, we now recast $g_m(z) = c_m\sum_n f_m(n)z^n$, which yields

\begin{equation}\label{eq:sol_fn}
\begin{split}
f_m(n) = (-1)^{m-n}\sum_{\ell=0}^n& \binom{N-m}{\ell}\binom{m}{n-\ell}\\
& \frac{\Gamma\left(a+\ell\right)}{\Gamma\left(a\right)}\frac{\Gamma\left(b\right)}{\Gamma\left(b+\ell\right)},
\end{split}
\end{equation}
with $a=m+\frac{\varepsilon}{\mu}$ and $b=1-N- \frac{\varepsilon}{\mu}$.\footnote{Note that the above formula is not defined when $\frac{\varepsilon}{\mu}$ is an integer because $b$ and $b+\ell$ are negative integers and therefore $\Gamma(b)$ and $\Gamma(b+\ell)$ are not defined. In that case, however, it is possible to use the Euler reflection formula by writing $\frac{\varepsilon}{\mu}=k+\delta$ with $k\in\mathbb{N}$, $0<\delta<1$ and then take $\delta \to 0$ to replace $\frac{\Gamma(b+\ell)}{\Gamma(b)}$ with $\frac{\Gamma\left(N+\frac{\varepsilon}{\mu}-\ell\right)}{\Gamma\left(N+\frac{\varepsilon}{\mu}\right)}\left(-1\right)^{\frac{\varepsilon}{\mu}}$.}  This describes the time-dependent solution up to the determination of the $c_m$ coefficients. We show our results for $m=0,1$ on Figure~\ref{fig:solutions}, and note that the agreement with the $N\to\infty$ results from~\cite{moran2020schrodinger} is remarkably good.

Noticing then that 
\begin{equation}\label{eq:asymptotic}
c_0 = \frac{1}{_2F_1\left(\frac{\varepsilon}{\mu}, -N; 1-N-\frac{\varepsilon}{\mu},1\right)} \underset{N\to\infty}{\approx} \frac{\Gamma\left(2\frac{\varepsilon}{\mu}\right)}{\Gamma\left(\frac{\varepsilon}{\mu}\right)}N^{-\frac{\varepsilon}{\mu}}
\end{equation}
and that $f_0(0)=1$, one has directly that the probability of having $n=0$ ants in the right-hand side food-source in the asymptotic regime behaves as $N^{-\frac{\varepsilon}{\mu}}$. In particular, if one does as in~\cite{kirman1993ants,moran2020schrodinger} and studies the asymptotic probability density corresponding of the \textit{fraction} of ants $n/N$ in the right-hand food source, multiplication by the Jacobian of the transformation means that the asymptotic density at $n/N=0$ behaves as $N^{1-\frac{\varepsilon}{\mu}}$. This corresponds to the behaviour of the density of the symmetric Beta distribution of parameter $\frac{\varepsilon}{\mu}$ at $0$, as given in Eq.~\eqref{eq:stationary}.

Nonetheless, it is possible to obtain the full time-dependent solution for a one-dimensional master equation such as~\eqref{eq:MEKirman} using the alternative method described in~\cite{smith2015general}, which is exact up to the determination of the eigenvalues of the transition rate matrix which we have already obtained above. Similar applications of this little-known method have been employed in several recent publications, for the solution of Brock and Durlauf's binary decision model \cite{holehouse2021non}, a solution to the Michaelis-Menten enzyme reaction \cite{holehouse2020stochastic}, and in solving the fast-switching autoregulatory genetic feedback loop with bursty gene expression \cite{jia2020dynamical}. Note that this method is very similar to the one described in \cite{ashcroft2016metastable,ashcroft2015mean}, although these publications use Laplace transforms instead of Cauchy's integral formula.

We shall now detail the essential steps from the method of \cite{smith2015general} in a generalised form that allows for multi-step reactions/events. For more rigorous details, see \cite{smith2015general}. We start again from the formal solution $\vec{P}(t) = e^{t\mathbf{M}}\vec{P}(0)$, which after using Cauchy's integral formula\footnote{This formula states that for any function $\vec{f}(\mathbf{M})=\frac{1}{2\pi \iu}\oint_\gamma \mathrm{d} z~ (z\mathbf{I}-\mathbf{M})^{-1}\vec{f}(z)$, where $\gamma$ is a contour containing the spectrum of $\mathbf{M}$.} reads

\begin{equation}\label{eq:cauchy_sol}
\vec{P}(t) = \frac{1}{2\pi \iu}\oint_\gamma \mathrm{d}z~e^{zt}(z\mathbf{I}-\mathbf{M})^{-1}\vec{P}(0).
\end{equation}

However, because $P(n,0)=\delta_{n,n_0}$ one can verify that $P(n,t)=[(z\mathbf{I}-\mathbf{M})^{-1}\vec{P}(0)]_n = [(z\mathbf{I}-\mathbf{M})^{-1}]_{n, n_0}$ (where $\mathbf{M}_{0,0}$ is the top-left-hand element of $\mathbf{M}$). We next use that for any invertible matrix $\mathbf{A}$, $\mathbf{A}^{-1} = \text{adj}(\mathbf{A})/\text{det}(\mathbf{A})$, where $\text{adj}(\mathbf{A})$ is the adjugate matrix of $\mathbf{A}$, or equivalently the transpose of the cofactor matrix. Defining $\mathbf{B}(z) = \text{adj}\left(z\mathbf{I}-\mathbf{M}\right)$ we therefore reach the following expression,

\begin{equation}
    P(n,t) =\frac{1}{2\pi i}\oint_\gamma \mathrm{d}z~ \frac{e^{zt}}{\prod_{i=0}^{N}(z+\lambda_i)}\mathbf{B}(z)_{n,n_0},
\end{equation}
where $\mathbf{B}(z)_{n,n_0}$ is a polynomial in $z$, as expected, and can be determined using standard methods~\cite{higham2008functions}, including a simple iterative formula for the case of tridiagonal $\mathbf{M}$, i.e., for a one-step birth death process~\cite{usmani1994inversion}, as is the case here. 

Evaluating the integral using Cauchy's residue theorem leads to the following expression,

\begin{align}\label{eq:Pnsol1D}
    P(n,t) = \sum_{m=0}^{N}\left\{e^{-\lambda_m t}\frac{\mathbf{B}(-\lambda_m)_{n,n_0}}{\prod_{j\neq m}(\lambda_j-\lambda_m)}\right\}.
\end{align}
where we now recognise the equivalence with the result obtained using generating functions,
\begin{equation}
    c_m f_m(n) =\frac{\mathbf{B}(-\lambda_m)_{n,n_0}}{\prod_{j\neq m}(\lambda_j-\lambda_m)}.
\end{equation}

To summarise our results, the generating function approach allowed us to obtain the eigenvalues $-\lambda_m$ and the functions $f_m$ describing the solution, up to the constants $c_m$ that depend on the initial state. The last approach, using Cauchy's integral formula, allowed us to obtain a more amenable expression that is easy to evaluate numerically provided we have the eigenvalues obtained previously. We apply these methods to simulate the time-evolution of the distribution $P(n,t)$ in the case of the symmetric model and the asymmetric generalisations considered below on Figure~\ref{fig:antplots}. Note that we validate our analytical results in Fig.~\ref{fig:antplots} against the stochastic simulation algorithm (SSA, \cite{gillespie2007stochastic}), a Monte Carlo method from which one can simulate exact stochastic trajectories describing master equations, for example Eq.~\eqref{eq:MEKirman} (Fig.~\ref{fig:antplots}, top plot).

\subsection{Extension to asymmetric sources}
\subsubsection{Asymmetric noise only}

The same analysis can be extended to the asymmetric ant model, studied in~\cite{Moran2021} to model the dynamics of fishing boats and in~\cite{alfarano005} to model agents trading in a financial market. This version of the model amounts to saying that the noise level $\varepsilon$ depends on whether an ant is in the left- or right-hand food source. In this case, Eq.~\eqref{eq:kirmanReac} becomes
\begin{equation}\label{eq:reac_asym}
L \xrightleftharpoons[n\varepsilon_2+n(N-n)\mu]{(N-n)\varepsilon_1 +n(N-n)\mu} R.
\end{equation}

The same analysis as above may be carried out in exactly the same way. After solving the eigenvalue problem using the characteristic function and imposing that it be a polynomial we find the following expression for the eigenvalues,
\begin{equation}\label{eq:asym_eigvals}
\lambda_m=m\left(\varepsilon_1+\varepsilon_2+(m-1)\mu\right), \quad m\in \llbracket 0; N\rrbracket,
\end{equation}
which is the same expression obtained in the continuum $N\to\infty$ version obtained in~\cite{Moran2021}.

Similarly, the modes $g_m(z)$ read
\begin{equation}\label{eq:gm_asym}
g_m(z) = \left(z-1\right)^{m}{}_2F_1\left(m+\frac{\varepsilon_1}{\mu},m-N;1-N-\frac{\varepsilon_2}{\mu},z\right),
\end{equation}
and the expressions for $f_m(n)$ are given by Eq.~\eqref{eq:sol_fn} but with $a=m+\frac{\varepsilon_1}{\mu}$ and $b=1-N- \frac{\varepsilon_2}{\mu}$.

\begin{figure}[tb]
    \centering
    \includegraphics[width=.5\textwidth]{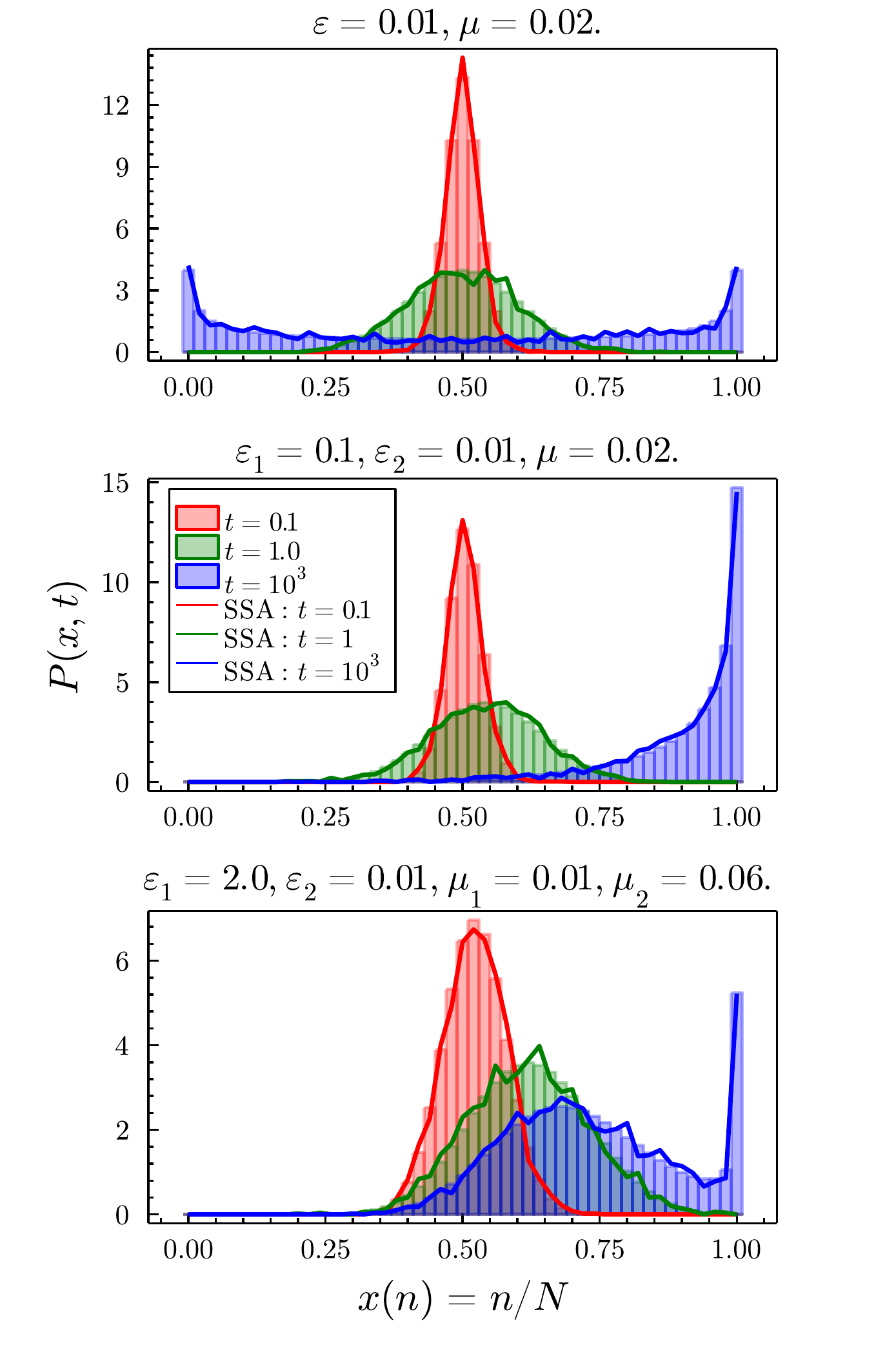}
    \caption{Plots showing the time evolution of ant rationality models under varying levels of asymmetry. In all plots the distributions are shown for $N=50$ agents and initial condition $x = 0.5$, with the histograms showing the analytic solution (from Eq.~\eqref{eq:Pnsol1D}) and solid lines showing ensemble distributions from 2500 simulations of the stochastic simulation algorithm (SSA) \cite{gillespie2007stochastic}. The top plot shows a time evolution for the completely symmetric ant model; the middle plot shows a time evolution for the asymmetric $\varepsilon$ model; and the bottom plot shows a time evolution for the entirely asymmetric case of $\varepsilon_1\neq\varepsilon_2$ and $\mu_1\neq\mu_2$. Clearly, as the model becomes more asymmetric more complex behaviours are possible.}
    \label{fig:antplots}
\end{figure}

Again in this case we find that the convergence rate is given by $\varepsilon_1 + \varepsilon_2$ and therefore does not depend on the imitation rate $\mu$ for any value of $N$. We verify our analytic solution in Fig.~\ref{fig:antplots} (middle plot) against the SSA.

\subsubsection{Full asymmetry}

The fully asymmetric case corresponds to a situation where the ants have a different imitation propensity depending on the food source they are currently in. Thus, Eq.~\eqref{eq:kirmanReac} now reads
\begin{equation}\label{eq:reac_fullasym}
L \xrightleftharpoons[n\varepsilon_2+n(N-n)\mu_2]{(N-n)\varepsilon_1 +n(N-n)\mu_1} R.
\end{equation}

The eigenvalue problem is now an ordinary differential equation with four regular singularities, and can therefore be solved via the Heun function~\cite[Sec.~31]{NIST:DLMF},

\begin{align}\label{eq:genGenFn}
    g_m(z) = H(a,q(\lambda_m);\alpha,\beta,\gamma,0;z),
\end{align}
where we define
\begin{equation}\label{eq:heundefs}
\begin{split}
    a &= \mu_2/\mu_1,\\
    q(\lambda_m) &= \frac{(\lambda_m-N\varepsilon_1)(N-1)}{\mu_1},\\
    \alpha &= -N,\\
    \beta &= \frac{(N-1)\varepsilon_1}{\mu_1},\\
    \gamma &= -(N-1)\left(1+\frac{\varepsilon_2}{\mu_2}\right).
    \end{split}
\end{equation}

We require again that this function be a polynomial of order $N$. We therefore write $H(a,q(\lambda_m);\alpha,\beta,\gamma,0;z)=\sum_{j=0}^\infty C_j z^j$, with the following recurrence relation (see ~\cite[Sec.~31.3]{NIST:DLMF}):
\begin{equation}
\begin{split}
    &C_0=1,\quad\alpha \gamma C_1 -q(\lambda_m(t))C_0=0,\\
    &R_j C_{j+1}-(Q_j+q(\lambda_m(t)))C_j+P_j C_{j-1} = 0,
    \end{split}
\end{equation}
with
\begin{equation}
\begin{split}
    R_j &= a(j+1)(j+\gamma),\\
    Q_j &= j((j-1+\gamma)(1+a)+1+\alpha+\beta-\gamma),\\
    P_j &= (j-1+\alpha)(j-1+\beta),
    \end{split}
\end{equation}
and naturally $C_j =0$ for $j>N$.

Setting $C_{N+1}=0$, this recurrence leads to an equation for $q(\lambda_m)$ using continued fractions. Writing $\frac{a_1}{b_1+\frac{a_2}{b_2+\ldots}}= \frac{a_1}{b_1+}\frac{a_2}{b_2+}\ldots$, we find
\begin{equation}\label{eq:contFrac}
    q(\lambda_m) =  \frac{R_0 P_1}{Q_1+q(\lambda_m)-}\frac{R_1P_2}{Q_2+q(\lambda_m)-}\dots \frac{R_{N-1}P_N}{Q_N+q(\lambda_m)},
\end{equation}
which then leads to a polynomial of order $N+1$ in $\lambda_m$ and therefore to the $N+1$ distinct eigenvalues. This case, therefore, does not lead to a situation where we can improve on, say, a direct diagonalisation of the transition rate matrix.

It is nonetheless possible to study the time evolution of all instances of the model numerically, as shown on Figure~\ref{fig:antplots} (bottom plot).

\section{Applications to other models}

A large class of binary decision models with interactions can be mapped onto birth/death processes. Indeed, if the dynamics is such that at every time step one or more agents change their mind from choice $A$ to $B$, then this can be rewritten as removing an agent of class $A$ from the population and replacing them with $B$. Thus it is possible to write a reaction scheme as we have done previously, write the master equation, find the corresponding differential equation for the generating function and solve using the methods we have shown. 

One of the methods we used was already used in~\cite{holehouse2021non} to solve the Brock and Durlauf model~\cite{brock2001discrete}. We further illustrate this by giving solutions to the voter and vacillating voter models.

\subsection{The voter model}
In the voter model~\cite{liggett1999stochastic} one is interested in the opinion dynamics of individuals who can vote for two distinct choices---voting for a left- or right-wing political party, say. We can again chose to label those choices by $L$ and $R$. 

The model imagines that the agents are embedded in a social network, and they only communicate with nearest neighbours. In the dynamics, with probability $p_d$ an agent is picked at random and their opinion becomes $L$ or $R$ with equal probability, or with probability $1-p_d$ a pair of neighbouring agents with opposite opinions $LR$ is chosen, and then one of the agents persuades the other into adopting their opinion, so that the new pair becomes $LL$ or $RR$ with equal probability. 

\begin{figure}[tb]
    \centering
    \includegraphics[width=.45\textwidth]{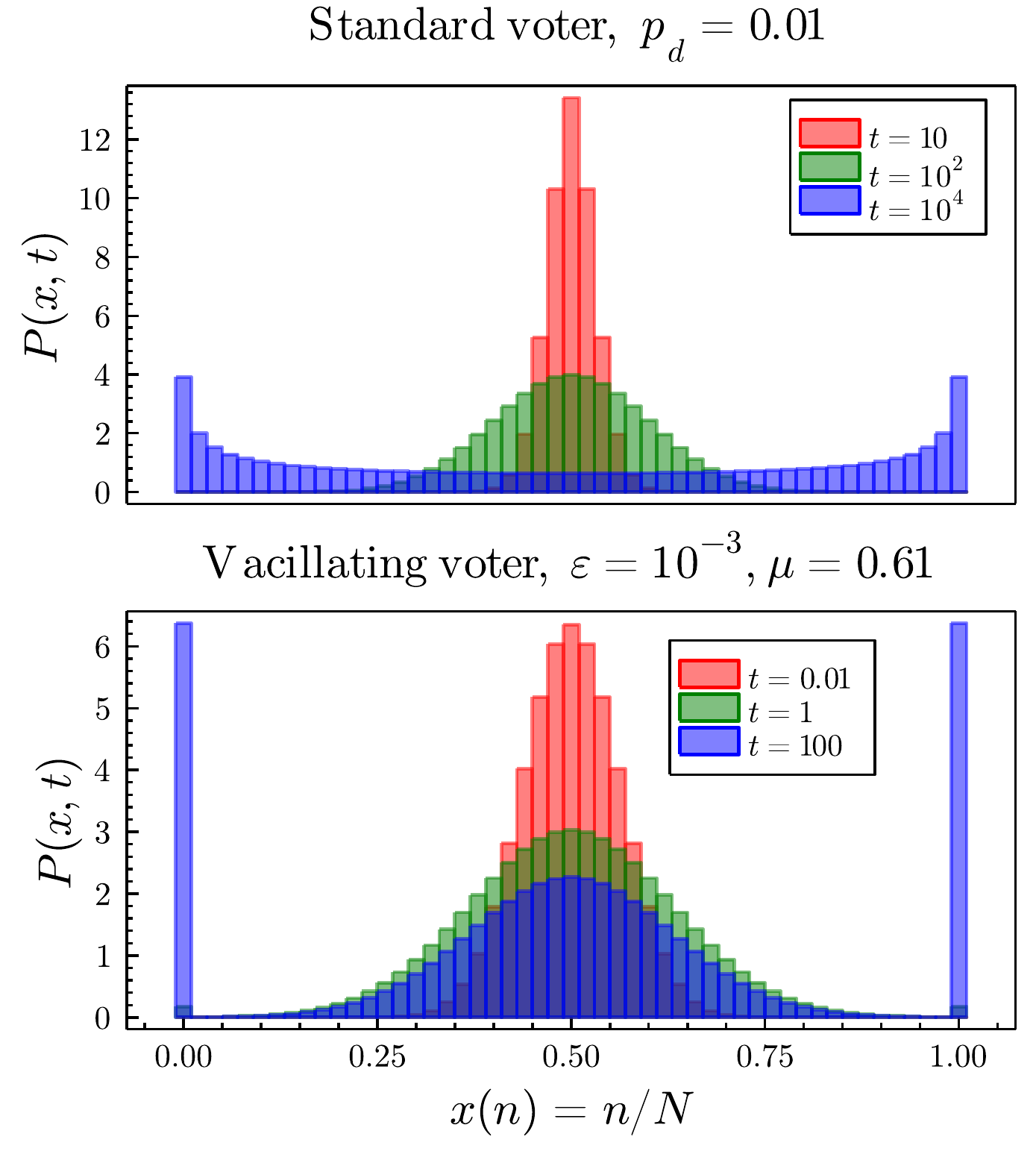}
    \caption{Plots showing the time evolution of the voter model (top) and the vacillating voter model (bottom) for $N=50$ agents and initial condition $x = 0.5$. Unlike the voter model, the vacillating voter model is capable of exhibiting steady state trimodality (seen here for $t=100$), brought on by the unsure nature of the voters.}
    \label{fig:voters}
\end{figure}

In this case, the model bears strong similarities with models for catalytic reactions between two different chemical species $L$ and $R$~\cite{fichthorn1989noise,considine1989comment,Krapivsky1992} that are embedded in a substrate onto which they have adsorbed. The interpretation is now that (i) with probability $p_d$ per unit time $L$ or $R$ desorb and are \textit{immediately} replaced (at equal probability) with $L$ or $R$, and (ii) with probability $1-p_d$ per unit time a nearest neighbour $LR$ pair react and desorb, and are \textit{immediately} replaced with $2L$s or $2R$s. The reaction scheme now reads 

\begin{equation}
        L\xrightleftharpoons[\frac{p_d x}{2}+\frac{(1-p_d)x(1-x)}{2}]{\frac{p_d(1-x)}{2}+\frac{(1-p_d)x(1-x)}{2}} R,
\end{equation}
where $x=n/N$ is the concentration of species $R$ in the substrate.

Clearly, this is just a special case of the ant recruitment model of Eq.~\eqref{eq:kirmanReac} in the special case where $\mu=(N-1)(1-2\varepsilon N)/2N^2$ and $\varepsilon< 1/2N$. Hence, its time-dependent dynamics is solved by the analyses above. We note that although it is a special case of the symmetric ant model it largely does share the models' phenomenology---showing the transition from monomodality to bimodality in the transient and steady state dynamics, albeit in a restricted section of the parameter space. A benefit of and Föllmer's ant model is that extensions towards higher degrees of asymmetry between the food sources are more easily implemented.

\subsection{The vacillating voter model}

Another version of said model is that of the \textit{vacillating} voter model~\cite{lambiotte2007dynamics}. This model extends the voter model to the case where agents are unsure of their opinion. The dynamics is as follows: every time-step, an agent $i$ with an opinion $S_i\in\{L, R\}$ is selected at random. With a probability $\propto \varepsilon$ the agent changes their mind randomly to the opposite choice, and with a probability $\propto \nu$ the agent then selects another agent $j\neq i$ at random. If $S_j\neq S_i$ then $i$'s opinion is updated as $S_i\leftarrow S_j$, but if instead $S_j = S_i $ and the agents already agree, then $i$ picks yet another random agent $k$. If $S_k\neq S_i$ then $S_i\leftarrow S_k$, and $i$ retains their original opinion if $S_k = S_i$. 

Again we may write the reaction scheme for this model,

\begin{equation}\label{eq:voter_reaction}
    L\xrightleftharpoons[\varepsilon n +\nu \frac{(N-n)n}{N-1}\left(1+\frac{n}{N-1}\right)]{\varepsilon (N-n) +\nu \frac{(N-n)n}{N-1}\left(1+\frac{N-n}{N-1}\right)}R,
\end{equation}
and apply the same reasoning as before. 

This now yields third-order ODEs for the $N$ generating function $g_m(z)$ (see Appendix~\ref{ap:voter}). Using the method of Frobenius, we write each $g_m(z)$ as a series and find the conditions for which it is a polynomial of degree $N+1$. This then allows to find a continued fraction expression that is satisfied by the eigenvalues $\lambda_m$, which again allows one to compute a full time-dependent solution using the resolvent relationship of Eq.~\eqref{eq:Pnsol1D}. Our numerical results are shown in Fig.~\ref{fig:voters}.

\section{Conclusion}

In this paper, we have provided an exact solution to the ant recruitment model. We have proved that we can recover the $N\to\infty$ results found through other methods in~\cite{moran2020schrodinger}, finding in particular that the stationary state is reached at an exponential rate of $2\varepsilon$, independently of $N$. 

We have also shown how our method can be extended to any binary decision model that can be mapped onto a one-step birth/death process. We have illustrated this with applications to the voter and the vacillating voter models. 

More interesting lines of research are however possible in the context of decision theory. For example, our method works very well for models that display microscopic reversibility, as the process of one or more agents changing their mind from $A$ to $B$ can also be reversed by the process. However, as highlighted in~\cite{bouchaudcrises}, more complicated interactions between agents can break microscopic reversibility (or break detailed balance, in physics parlance) and possibly lead to more interesting phenomena. 

A promising way to introduce this is through explicit path-dependency in the agents' decision-making process. Such effects have been studied through the inclusion of memory effects in utility functions~\cite{Moranhabit,harris_peak,Mitsokapas2021}, showing interesting effects such as ageing or memory induced condensation. There remains, however, to see how such memory effects could arise naturally from interactions. It should be noted, as highlighted in~\cite{bouchaud2021selffulfilling}, that under a timescale of order $1/2\varepsilon$ the model we have described here is not ergodic: in the high imitation regime one may think that one of the two choices is optimal because it has been made all the time so far, but this may be only because under that timescale the ants are self-consistently ``trapped'' in one given choice, and one has not had the time to observe a full collective switch.

Other exciting results in decision theory can be obtained with random interactions. Agents interacting through random games are already known to produce very rich dynamics, in particular because multiple Nash equilibria emerge as the games become more complex~\cite{wiese2020frequency} and because minute details such as the order in which players update their actions has an impact on the existence of an equilibrium~\cite{heinrich2021bestresponse}.

Similarly, if the agents influence each other via a network with random topology and random weights then one can expect the dynamics to be similar to that of a spin glass, and therefore to display a rich phase diagram and non-intuitive dynamical behaviour. Progress in this direction has been made e.g. in~\cite{baron}, whose dynamics strongly resembles those describing glassy population dynamics in large ecosystems~\cite{altieri,roy2019endogenous} and where we can expect path dependency to emerge naturally.

By further studying the dynamics of economic toy models, such as those considered in this paper, we can further expand the library of qualitative transient behaviours one expects to see in more complex models. Such qualitative behaviours are highly valuable, since they allow us to reframe the behaviours of real economic data, and parse them with respect to well understood behaviours from simpler models.

\section*{Acknowledgements}
J.H.~would like to thank Kaan \"Ocal for useful conversations in the inception of this paper. J.M.~thanks Jean-Philippe Bouchaud, Michael Benzaquen and Théo Dessertaine for numerous conversations about behavioural modelling. J.H. was supported by an BBSRC EASTBIO Ph.D. studentship. We thank Ramon Grima and Sidney Redner for useful comments on the original manuscript.



\bibliographystyle{naturemag.bst}
\bibliography{main}

\pagebreak
\clearpage
\widetext
\begin{center}
\textbf{\large Appendix: {\it Exact time-dependent dynamics of discrete binary choice models}}
\end{center}
\setcounter{equation}{0}
\setcounter{figure}{0}
\setcounter{table}{0}
\setcounter{footnote}{0}
\setcounter{page}{1}
\makeatletter
\renewcommand{\theequation}{S\arabic{equation}}
\renewcommand{\thefigure}{S\arabic{figure}}
\renewcommand{\bibnumfmt}[1]{[S#1]}
\renewcommand{\citenumfont}[1]{S#1}
\appendix

\section{Calculation of $c_m$ from Sturm--Liouville theory}\label{ap:sl_theory}
In this Appendix we complete the specification of the generating function from Eq.~\eqref{eq:gzkirman} in the main text, explaining how the coefficients $c_m$ may be computed from an initial condition.
Aside from the first coefficient that corresponds to the weight on the steady-state generating function, given by $c_0=1/g_{1,0}(1)$, the other coefficients are determined from the initial condition. We now look to determine the non-zero coefficients $c_{m\geq 1}$ from Sturm--Liouville theory \cite{zettl2010sturm}.

Consider a second-order linear ODE of the same type as Eq.~\eqref{eq:GFE} in the main text,
\begin{align}
    \left[\beta_1(z)\partial_z^2+\beta_2(z)\partial_z+\beta_3(z)\right]f(z,t) = \partial_t f(z,t),
\end{align}
which can be solved using separation of variables to obtain the general solution
\begin{align}
    f(z,t)=\sum_m b_m F_m(z)e^{-\lambda_m t},
\end{align}
where each $F_m(z)$ is a linearly-independent eigenfunction, i.e. a solution of,
\begin{align}
    \hat{O}F_m(z) \equiv \beta_1(z)F_m''(z)+\beta_2(z)F_m'(z)+\beta_3(z)F_m(z) = -\lambda_m F_m(z),
\end{align}
and $-\lambda_m$ are the eigenvalues of $\hat{O}$. 

Sturm--Liouville theory states that the eigenfunctions will form an orthogonal basis under the $w$-weighted inner product in the Hilbert space $L^2([a,b],w(z)dz)$ denoted,
\begin{align}
    \langle F_n(z),F_m(z)\rangle\equiv \int_{a}^b F_n(z)F_m(z)w(z)~\mathrm{d}z\propto \delta_{n,m},
\end{align}
where $w(z)$ is given by,
\begin{align}
    w(z)=\frac{1}{\beta_1(z)}e^{\int \frac{\beta_2(z)}{\beta_1(z)}~\mathrm{d}z}.
\end{align}
This orthogonality property then allows one to find the coefficient $b_m$ with respect to projections onto the initial state,
\begin{align}
    b_m = \frac{\langle F_m(z),q(z) \rangle}{\langle F_m(z),F_m(z) \rangle}.
\end{align}
In our case, from Eq.~\eqref{eq:GFE} we find that 
\begin{equation}
     w(z) = (1-z)^{\frac{2\varepsilon}{\mu}-1}z^{-(N+\frac{\varepsilon}{\mu})}
 \end{equation} 
and $[a,b]=[-1,1]$, as it is the region over which the generating function is defined.

One can then find the coefficients $c_m$ as,
\begin{align}
    c_m = \frac{\langle g_m(z),z^{n_0} \rangle}{\langle g_m(z),g_m(z) \rangle},
\end{align}
where the initial number of ants on the right-hand source is $n_0$. This completes the specification of the generating function solution which is now simply given by,
\begin{align}
    G(z,t) = \sum_{m=0}^{N}c_m(z-1)^{m}{}_2F_1\left(m+\varepsilon/\mu,m-N;1-N-\varepsilon/\mu,  z\right)e^{-m\left(2\varepsilon+(m-1)\mu\right) t}.
\end{align}

\section{Solution to the vacillating voter model}\label{ap:voter}

Starting from the reaction scheme~\eqref{eq:voter_reaction}, one obtains the following ordinary differential equation for the functions $g_m(z)$:

\begin{equation}\label{eq:voter_ODE}
\begin{split}
&\nu  z^2(z+1)(z-1) g_m'''(z)\\
&+ \nu z(z-1)(2+4z-3Nz)g_m''(z)\\
&- (N-1)(z-1)((N-1)(z+1)\varepsilon+\nu(N+2z(1-N))) g_m'(z)\\
&+ (N-1)(\lambda_m +N(z-1)\varepsilon)g_m(z)=0.
\end{split}
\end{equation}
We next obtain a recursion relation for the coefficients $C_{j}$, as

\begin{equation}\label{eq:rec}
\begin{split}
&C_0=1,\quad(N-1)((N-1)\varepsilon+N\nu) C_1 -q(\lambda_m)C_0=0,\\
&R_j C_{j+1}-(Q_j+q(\lambda_m(t)))C_j+P_j C_{j-1} = 0,
\end{split}
\end{equation}
with the condition that $C_{N+1}=0$, and where we write
\begin{equation}\label{eq:rec_Defs}
\begin{split}
q(\lambda_m) =& (N-1)(\varepsilon N-\lambda_m),\\
R_j =& (j+1) \bigg(j \bigg(-j^2+j-2\bigg) \nu\\
& +\nu  N (N-1)+(N-1)^2 \varepsilon \bigg),\\
Q_j =& -j \nu  (3 N-2) (N-j),\\
P_j =& (j-1) \nu  (j-2 N) (j-N-1)\\
&-(N-1) \varepsilon  ((j-2) N-j+1).
\end{split}
\end{equation}
We obtain again a continued fraction relation that determines the eigenvalues $\lambda_m$,
\begin{equation}\label{eq:contFrac2}
\begin{split}
q(\lambda_m(t)) =
  \frac{R_0 P_1}{Q_1+q(\lambda_m(t))-}\frac{R_1P_2}{Q_2+q(\lambda_m(t))-}\dots \frac{R_{N-1}P_N}{Q_N+q(\lambda_m(t))}.
\end{split}
\end{equation}
Calculating $\lambda_m$ from this relation, the full time-dependent solution is given using the resolvent relationship in Eq.~\eqref{eq:Pnsol1D}.

\end{document}